\documentclass[12pt,preprint]{aastex}

\slugcomment{To appear in the {\it Astrophysical Journal Supplement Series}}

\begin{document}

\title{The VLA 1.4~GHz Survey of the\\
Extended Chandra Deep Field South: First Data Release}

\author{Neal A. Miller\altaffilmark{1,2}} 
\email{nmiller@pha.jhu.edu}

\author{Edward B. Fomalont\altaffilmark{3}}

\author{Kenneth I. Kellermann\altaffilmark{3}}

\author{Vincenzo Mainieri\altaffilmark{4}}

\author{Colin Norman\altaffilmark{2}}

\author{Paolo Padovani\altaffilmark{4}}

\author{Piero Rosati\altaffilmark{4}}

\author{Paolo Tozzi\altaffilmark{5}}

\altaffiltext{1}{Jansky Fellow of the National Radio Astronomy Observatory. The National Radio Astronomy Observatory is a facility of the National Science Foundation operated under cooperative agreement by Associated Universities, Inc.}
\altaffiltext{2}{Department of Physics and Astronomy, Johns Hopkins University, 3400 N. Charles Street, Baltimore, MD 21218}
\altaffiltext{3}{National Radio Astronomy Observatory, 520 Edgemont Road, Charlottesville, VA 22903-2475}
\altaffiltext{4}{European Organisation for Astronomical Research in the Southern Hemisphere, Karl-Schwarschild-Strasse 2, D-85748, Garching bei Munchen, Germany}
\altaffiltext{5}{INAF Osservatorio Astronomico di Trieste, via G.B. Tiepolo 11, I-34131, Trieste, Italy}

\begin{abstract} 
We have observed the Extended {\it Chandra} Deep Field South (E-CDF-S) using a mosaic of six deep Very Large Array (VLA) pointings at 1.4~GHz. In this paper, we present the survey strategy, description of the observations, and the first data release. The observations were performed during June through September of 2007 and included from 15 to 17 ``classic'' VLA antennas and 6 to 11 that had been retrofitted for the Expanded VLA (EVLA). The first data release consists of a 34\farcm1$\times$34\farcm1 image and the attendant source catalog. The image achieves an rms sensitivity of 6.4 $\mu$Jy per 2\farcs8$\times$1\farcs6 beam in its deepest regions, with a typical sensitivity of 8 $\mu$Jy. The catalog is conservative in that it only lists sources with peak flux densities greater than seven times the local rms noise, yet it still contains 464 sources. Nineteen of these are complex sources consisting of multiple components. Cross matching of the catalog to prior surveys of the E-CDF-S confirms the linearity of the flux density calibration, albeit with a slight possible offset (a few percent) in scale. Improvements to the data reduction and source catalog are ongoing, and we intend to produce a second data release in January 2009.
\end{abstract}

\section{Introduction}\label{sec-introduction}

The {\it Chandra} Deep Field South \citep[CDF-S;][]{giacconi2002} is proving to be one of the more important deep fields for multiwavelength study of galaxy evolution. Originally consisting of a 1 Msec ACIS observation of a single field (about $16^\prime \times 16^\prime$), its areal coverage was expanded through four adjacent 250 ksec observations \citep[the Extended CDF-S, or E-CDF-S;][]{lehmer2005} and the CDF-S itself was recently observed for an additional 1 Msec of director's discretionary time. The {\it Hubble Space Telescope} and {\it Spitzer Space Telescope} have also targeted the CDF-S and E-CDF-S. The southern portion of the Great Observatories Origins Deep Survey \citep[GOODS;][]{dickinson2003} coincides with the CDF-S, and the {\it Hubble} Ultradeep Field \citep[HUDF;][]{beckwith2006} is also situated there. The larger E-CDF-S area has also been observed by {\it Hubble} and {\it Spitzer} through the Galaxy Evolution from Morphologies and SEDs program \citep[GEMS;][]{rix2004} and the {\it Spitzer} Wide-Area Infrared Survey \citep[SWIRE;][]{lonsdale2003}, respectively. In addition to the Great Observatories, satellite observations of the full E-CDF-S area have been made by the Galaxy Evolution Explorer \citep[{\it GALEX};][]{martin2005} in the ultraviolet as a target of both its Ultra-Deep Imaging Survey and Deep Spectroscopic Survey. Similarly, the field is rich in ground-based data, particularly when providing the redshifts that are necessary to cosmological studies. The ``Classifying Objects by Medium-Band Observations'' 17-filter survey \citep[COMBO-17;][]{wolf2004} spans nearly the entire E-CDF-S area, with the numerous passbands producing excellent photometric redshifts (accuracies of about 2\% or better in $\delta_z / (1+z)$ to $R \sim 22$). Spectroscopic redshifts have also been collected for thousands of galaxies \citep[e.g.,][]{szokoly2004,lefevre2004,mignoli2005,vanzella2005,vanzella2006,vanzella2008,ravikumar2007}.

A pair of programs have obtained deep 1.4~GHz radio observations and presented subsequent analysis in the CDF-S and E-CDF-S regions. Using the Australia Telescope Compact Array (ATCA), \citet{norris2006} examined a 3.7 deg$^2$ area including the E-CDF-S and its associated SWIRE region. They adopted a mosaic strategy consisting of 28 total pointings, with the seven pointings centered on the CDF-S (and hence the southern GOODS field) receiving deeper integrations and being somewhat more tightly spaced than the remainder of the mosaic. These more targeted observations are described in Koekemoer et al. (in preparation), and reach an rms sensitivity of about 14 $\mu$Jy per $17^{\prime\prime} \times 7^{\prime\prime}$ beam. \citet[][ hereafter A2006]{afonso2006} details the 64 ATCA radio detections within the GOODS area, finding optical and X-ray identifications and source characterizations for 58 of them. The connection between radio and X-ray sources was explored over the full E-CDF-S area in \citet{rovilos2007}, where it was found that the A2006 radio observations detected $14\%$ of the X-ray sources from the CDF-S and $9\%$ of those in the shallower E-CDF-S area. Interestingly, the 24$\mu$m {\it Spitzer} data suggested that over half of the radio/X-ray AGN exhibited evidence for concurrent star formation but that contrary to prior results \citep[e.g.,][]{bauer2002} there did not appear to be a link between faint radio sources and obscured X-ray AGN. AGN feedback models \citep[e.g.,][]{hopkins2005} postulate such a link, wherein a nuclear starburst obscures the central AGN and fuels it until outflows extinguish the starburst and allow the AGN to shine as an unobscured source.

\citet[][ hereafter K2008]{kellermann2008} observed the E-CDF-S using the National Radio Astronomy Observatory's (NRAO) Very Large Array (VLA). A single pointing at 1.4~GHz centered on the CDF-S produced an rms sensitivity of 8.5 $\mu$Jy per 3\farcs5 beam. A set of four pointings at 4.86~GHz yielded about the same sensitivity and resolution, thereby providing spectral index information for many of the 266 sources detected at 1.4~GHz. Additional observations at 4.86~GHz have been made and the improved 4.86~GHz image should have an rms sensitivity of about 6.5 $\mu$Jy, thereby increasing the number of sources with reliable spectral index measurements. \citet{mainieri2008} matched the radio sources to deep optical and near-IR data, associating 95\% (254/266) of the radio sources with optical/near-IR objects. The majority of these (191) had either photometric or spectroscopic redshift information. Along with Padovani et al. \citep[in preparation, and see][]{padovani2007}, they evaluated the composition of the fainter radio sources responsible for the flattening of the Euclidean normalized radio source counts below about 300 $\mu$Jy. Although sources powered by star formation become greater contributers to the source counts at these levels, AGN continue to be an important population. The X-ray spectral properties of the radio sources will be discussed in Tozzi et al. (in preparation).

However, each program has some limitations. The ATCA survey is hampered by its lower resolution; for a source at $z=1$, 1\arcsec{} equals 8 kpc (for $H_0 = 70$ km s$^{-1}$ Mpc$^{-1}$, $\Omega_m$ = 0.3, $\Omega_\Lambda = 0.7$) and hence the ATCA beam is about 135 kpc $\times$ 55 kpc. A 5$\sigma$-detected radio source at this redshift would have a rest-frame $L_{1.4~GHz} = 3 \times 10^{23}$ W Hz$^{-1}$, assuming a spectral index of 0.7 ($S_\nu \propto \nu^{-0.7}$). Locally, this 1.4~GHz luminosity corresponds to only the brightest radio-emitting galaxies, the vast majority of which are powerful AGN with radio jets and lobes \citep[e.g., see the RLF of ][]{condon2002}. The K2008 VLA survey has very good resolution (about 28 kpc at $z=1$) and is about twice as sensitive, but because it was performed using a single pointing its sensitivity declines radially from the CDF-S center. This means that outside the formal CDF-S area the ATCA survey is more sensitive.

We report here on a new VLA program that provides deep, high resolution 1.4~GHz imaging across the full E-CDF-S. Recognizing the importance of this field to the astronomical community and the rapid rate at which new multiwavelength data are becoming available, we are following a tiered approach to the release of the data. In this paper we present the first data release, consisting of the survey description, methods, and a best ``first pass'' image and catalog. This covers a $34^\prime \times 34^\prime$ region including the full E-CDF-S at a typical rms sensitivity of 8 $\mu$Jy per 2\farcs8$\times$1\farcs6 beam. We anticipate that with more refined imaging techniques the survey will ultimately achieve an rms sensitivity of around 7 $\mu$Jy per beam. A second data release consisting of these improved images and a more complete catalog is intended for release in January 2009.

The end goal of 7 $\mu$Jy rms sensitivity was determined based on the existing and soon-to-be completed multiwavelength surveys. The prior radio surveys indicate that this sensitivity will yield 3$\sigma$ or greater detections of the majority of the 762 X-ray sources in the \citet{lehmer2005} E-CDF-S catalog. In addition, a new {\it Spitzer} Legacy survey (``FIDEL,'' for Far-Infrared Deep Extragalactic Legacy survey\footnote{PI M.\ Dickinson, see {\ttfamily http://www.noao.edu/noao/fidel/}}) will produce the deepest 70$\mu$m images obtained by {\it Spitzer} (0.5 mJy rms) across nearly the entire E-CDF-S coverage area. The target rms sensitivity of 7 $\mu$Jy at 1.4~GHz for the VLA data is a direct match to the FIDEL sensitivity for a $z=1$ star-forming galaxy should it lie on the locally-observed FIR-radio correlation \citep[e.g.,][]{condon1991,yun2001}. Based on the extrapolation of an existing 70$\mu$m program reaching the depth of FIDEL \citep{frayer2006}, about 100 star-forming galaxies with $z\geq1$ should be detected by FIDEL. Should the FIR-radio correlation hold at such redshifts, nearly all should be detected by these new radio observations.

The paper is organized as follows. In Section \ref{sec-observations} we detail the observational strategy and schedule. The data reduction is described in Section \ref{sec-reductions}, and the initial source catalog in Section \ref{sec-catalog}. Each of these sections notes some of the considerations that should be kept in mind for users of the data in this first release. Section \ref{sec-future} provides a summary of the improvements that are planned for the second data release, and we summarize the key points of the current release in Section \ref{sec-summary}.

\section{Observational Strategy}\label{sec-observations}

\subsection{General Considerations}

\begin{figure}
\figurenum{1}
\epsscale{0.9}
\plotone{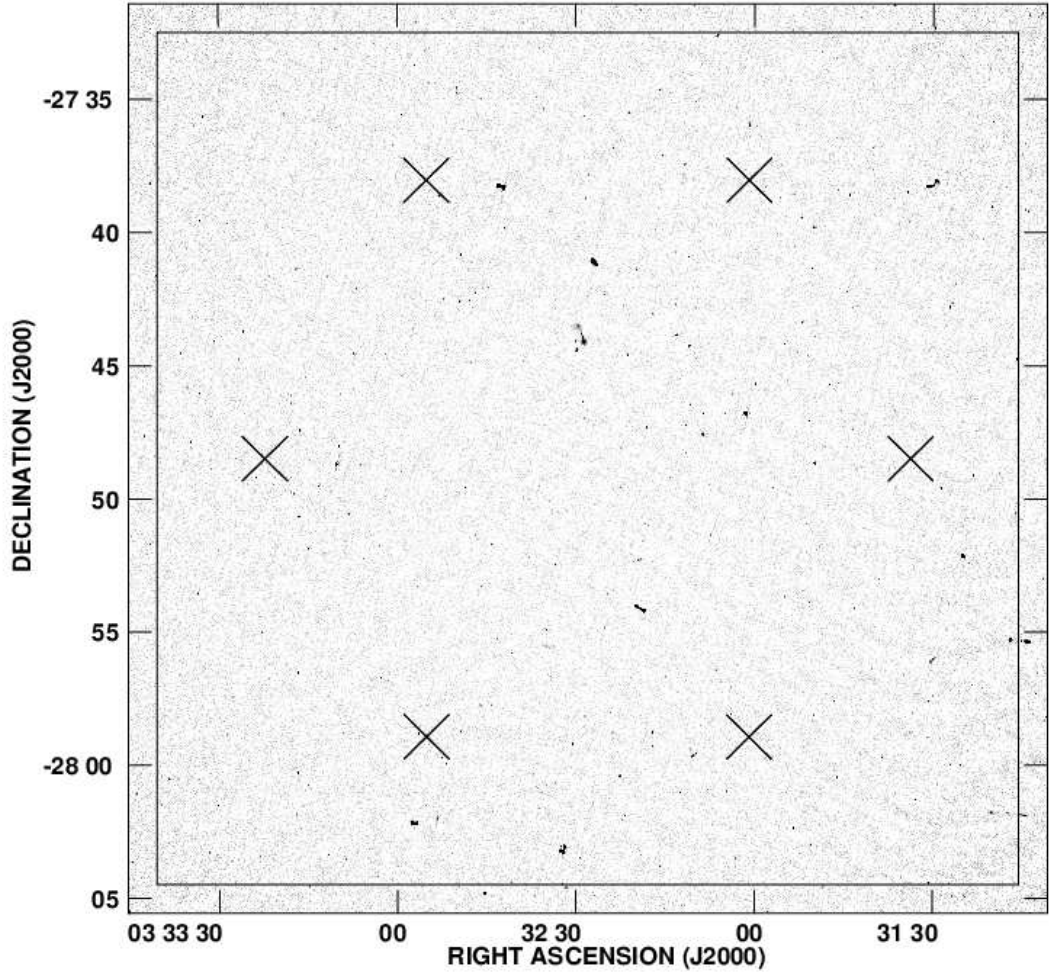}
\caption{Greyscale of first data release mosaic image. The six pointing centers are indicated by crosses, and the $32^\prime \times 32^\prime$ area of the E-CDF-S X-ray data is shown as the large square.\label{fig-msc}}
\end{figure}

\begin{figure}
\figurenum{2}
\epsscale{0.9}
\plotone{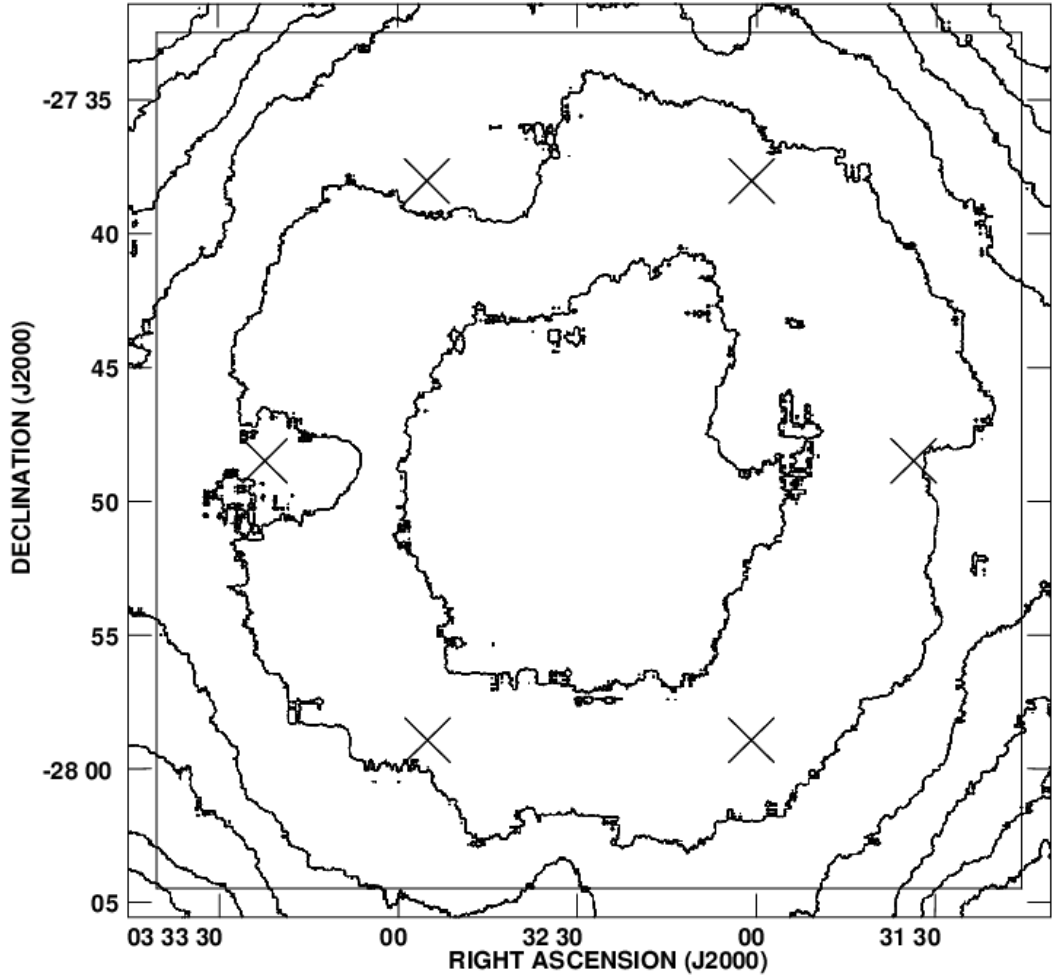}
\caption{Contours of constant rms sensitivity for first data release mosaic image. The innermost contour is at 7 $\mu$Jy beam$^{-1}$, with successive contours at increments of 1 $\mu$Jy beam$^{-1}$. As in Figure \ref{fig-msc}, the six pointing centers and the $32^\prime \times 32^\prime$ area of the E-CDF-S X-ray data are also shown.\label{fig-rms}}
\end{figure}

The primary beam of the VLA is about 31\arcmin{} full-width at half maximum, and smearing effects due to the observational bandwidth and integration time decrease this further. It is therefore impossible to provide near-uniform sensitivity across the full $32^\prime \times 32^\prime$ area of the E-CDF-S through a single pointing. We consequently adopted the standard practice of observing a hexagonal ring of six pointings with each one spaced 12\arcmin{} from its nearest neighbor. The center of the ring was the approximate center of the E-CDF-S, taken to be (J2000) 3$^{\mbox{{\scriptsize h}}}$32$^{\mbox{{\scriptsize m}}}$28\fs0 $-$27$^\circ$48\arcmin30\farcs0. This is the field center used for the K2008 observations, but was not the field center for a pointing in the current program. However, because it is well within the half-power point of each of the six individual pointings it still corresponds to the most sensitive region of the final image. The coordinates for each of the six pointings are included in Table \ref{tbl-obs} and depicted in Figures \ref{fig-msc} and \ref{fig-rms}.

Wide-field imaging with the VLA at 1.4~GHz is subject to several challenges, which may generally be attributed to the large size of the primary beam and the high density of sources at 1.4~GHz. Sources distant from the field center are distorted due to the observed bandwidth, causing them to smear in the radial direction. This limits the sensitivity of an image even at the field center, since more distant sources and their sidelobes are smeared and hence not easily cleaned. The problem is usually circumvented by dividing the full bandwidth into multiple smaller-sized channels, each of which may be imaged individually and then averaged. For 1.4~GHz observations with the VLA this requires an exchange of the usual 50~MHz continuum band for a 25~MHz band which can be further divided into channels. The loss of net bandwidth reduces the sensitivity at the field center, but for deep observations this loss is more than offset by the reduced bandwidth smearing and hence better sensitivity distant from the field center. Thus, the E-CDF-S observations were performed in this mode, consisting of seven 3.125~MHz channels at each of two center frequencies, 1.365~GHz and 1.435~GHz. Left and right circular polarizations were collected for each set, resulting in a total of 28 channels. The 3.125~MHz channels limit the bandwidth smearing for an individual pointing to an amount approximately equal to the resolution for sources 15\arcmin{} from the field center of that pointing. However, with the current VLA correlator limitations the use of channels a factor of two narrower would also decrease the net bandwidth by a factor of two.

\subsection{Scheduling and Implementation of Observations}

The program was awarded 240 hours of time as a result of the NRAO ``Large'' proposal solicitation. The program code, useful for obtaining the raw data through the VLA Archive, is AM889. The E-CDF-S is observable by the VLA for only about seven hours a day on account of its southern declination, and this is compounded by the VLA's increasing system temperature at 1.4~GHz as the elevation of the target decreases. Consequently, the observations were scheduled in 5-hour blocks centered at 3:30 LST, meaning the E-CDF-S was observed at elevations ranging from about 20 to 28 degrees. Thus, 48 individual days of observation were required to obtain the net allotment of 240 hours. The program ran from 03 June 2007 through 23 September 2007, with most of the observations occurring between the middle of July and early September (see Table \ref{tbl-obs}). With the exception of the final observation date the array was in its largest configuration, the A array. Due to some time lost due to unforeseen events (e.g., power outages at the VLA site, minor glitches related to the transition to EVLA) and rescheduled later, observations were actually scheduled on a total of 52 days.

We elected to devote each individual observation date (hereafter often referred to as a ``track'') to a single pointing of the six-pointing mosaic. The main reason for this was pragmatic, in that it lessens the complexity of the data reduction (see below). However, it should also prove useful to future studies of possible time variability of faint radio sources since observing a single pointing per track implies that a reasonably deep limit ($\sim30$ $\mu$Jy rms) is achieved on eight separate occasions. We sequentially cycled through the pointings, meaning each one was observed roughly once every ten days.

The same calibrators and general observing sequence were followed for each track. Flux calibration was achieved through observations of 3C48, with phase and bandpass calibration provided by the source J0340-213. This calibrator lies less than $7^\circ$ from the E-CDF-S and has a position known to an accuracy of better than 0.002 arcseconds from the VLBA Calibrator Survey \citep{beasley2002}. We alternated scans of J0340-213 and the target pointing with a roughly half hour cycle interval, with approximate time on source of 2.5 and 26.5 minutes, respectively. 3C48 was usually observed twice per track. The integration time per visibility was set at 3.3s in order to minimize the effect of time-averaging on the visibilities while keeping the data volume reasonable.

\section{Data Reduction}\label{sec-reductions}

\subsection{General Procedures}

A standard procedure was followed for each observing track. We used the 31DEC07 version of AIPS, updated regularly throughout the observing program. The raw (u,v) data were loaded, including back end system temperatures to be used for data weighting. We elected to perform our own flagging of all data since the new VLA online control software was known to occasionally flag good data and not all the system flags had been implemented yet for the antennas that had been retrofitted for the EVLA. The (u,v) data for J0340-213 were inspected and obvious time periods of bad data were removed (for example, the beginnings of scans where the array was not yet on source). The use of multiple channels requires proper calibration of the individual bandpass responses, so the J0340-213 data were used for this purpose. Because this source has some extended structure, this calibration only included baselines larger than 15 k$\lambda$. The derived bandpass calibration was then applied for subsequent data editing, beginning with 3C48. Whenever possible we used the scan of 3C48 made at elevations comparable to the E-CDF-S and J0340-213 for flux density calibration, ignoring the other scan which was made at higher elevation. The flux density of 3C48 was taken to be 16.38 Jy and 15.75 Jy at 1.365~GHz and 1.435~GHz, respectively, calculated according to \citet{baars1977}. This flux density scale was bootstrapped to J0340-213, which was then used to calibrate the gains and phases of the E-CDF-S data through linear interpolation. J0340-213 did appear to vary over the three months of observations, with its derived flux density declining steadily from about 1 Jy in June to 0.9 Jy in early August, then rising back to 1 Jy over the next two weeks where it remained until the observations concluded in September. We typically applied the derived calibrations and inspected all the 3C48 and J0340-213 data again, compiling additional flags as necessary and regenerating the calibrations if necessary. Once good calibration was achieved, the target E-CDF-S data were edited. The beginnings of each scan were removed, using the same duration found for the beginnings of the J0340-213 scans. Obviously bad data, usually restricted to a single antenna for fixed time ranges, were excised. In most cases these were directly corroborated by the observing logs for that date and included explanations such as an antenna receiving standard maintenance or suffering a known mechanical problem. Editing of the E-CDF-S data sometimes also included excision of radio frequency interference (RFI), as discussed further below (\ref{sec-challenges}).

\begin{figure}
\figurenum{3}
\epsscale{0.9}
\rotatebox{270}{\plotone{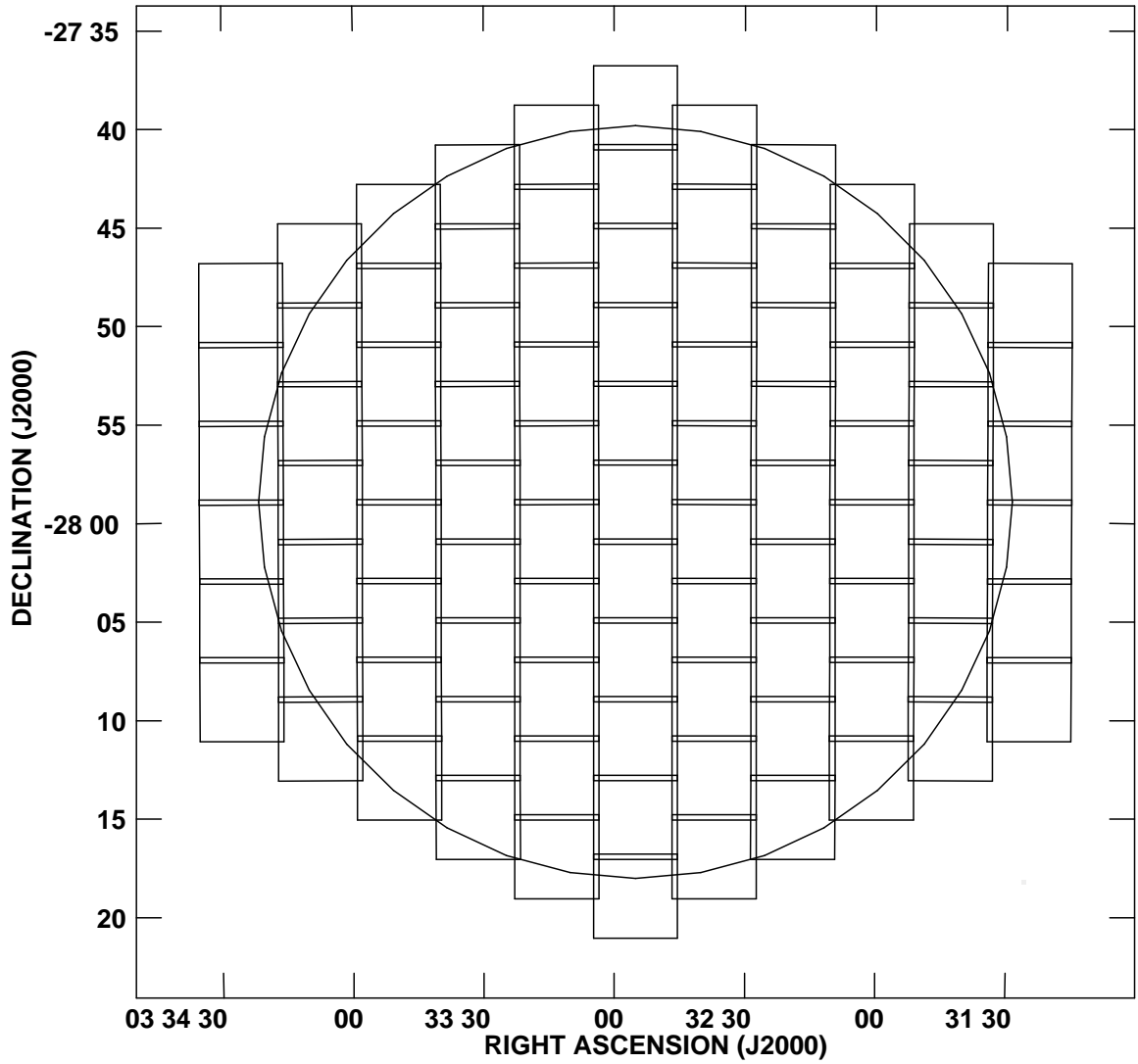}}
\caption{Arrangement of facets comprising the ``flys-eye'' for pointing 6. The large circle indicates the one-third power point of the primary beam.\label{fig-flyseye}}
\end{figure}

The target E-CDF-S data for that date's pointing were then imaged using the AIPS IMAGR task. First, we used the data to create a very large ($4^\circ$ across) image at low resolution. Upon this was overlaid a grid of 91 overlapping $256^{\prime\prime} \times 256^{\prime\prime}$ fields, arranged in a ``fly's-eye'' pattern (see Figure \ref{fig-flyseye}; a central field, surrounded by a ring of six fields, with this ring surrounded by a second ring of 12 fields, on out to a fifth ring, in total approximating a 22\arcmin{} radius circle). This fly's-eye represents the fields to be imaged at the full resolution of the data, meaning the large low-resolution map could be used to identify sources outside this area. Typically about 25 such ``flanking'' sources were noted, and their positions from the NVSS \citep{condon1998} were used as the field centers for subsequent imaging. 

The resulting $\sim115$ fields were then imaged using the multi-facet imaging capability of IMAGR. A pixel size of 0\farcs5 was used, and each field (facet) in the fly's-eye was $512 \times 512$ pixels (hence the $256^{\prime\prime} \times 256^{\prime\prime}$ fields in the fly's-eye). The flanking fields were usually smaller, at 128 pixels $\times$ 128 pixels, sufficient to image the bright source and hence prevent its sidelobes from adversely effecting the sensitivity within the fly's-eye facets. The use of multiple facets greatly alleviates image distortions due to sky curvature (the ``3D effect'') by shifting the (u,v) data for that facet to the tangent point at its center. IMAGR properly handles sources that are present in multiple facets, with the full flux density associated with a given coordinate being restored to each facet which includes that coordinate. After preliminary images of the facets were generated, we proceeded to box all identifiable sources. These boxes represent the allowable regions for IMAGR to search for sources to clean, thereby speeding subsequent imaging runs and more importantly preventing clean bias. Our imaging did apply a slight Gaussian taper to the (u,v) data along with a data weighting intermediate between pure uniform and pure natural weighting. This produced a well-behaved (nearly Gaussian) synthesized beam of 2\farcs8 $\times$ 1\farcs6 having a position angle near zero (i.e., north-south). Finally, several of the brightest sources within the fly's-eye coverage were given their own small facets for subsequent imaging. This improves their cleaning in the same manner as noted for imaging multiple facets.

Additional calibration and editing of the data were then performed. The images themselves can serve as the model input for further improvement of calibrated gains and phases, a procedure known as self calibration \citep[e.g.,][]{cornwell1989}. This helps to remove residual phase variation caused by variation in the troposphere and ionosphere above the telescopes in the array. For the data corresponding to an individual day's observations we used a four-minute interval for the self-calibration and did not alter the amplitudes. The self-calibrated data were again imaged, and further minor edits were performed by subtracting the clean components of the resulting images from the self-calibrated (u,v) data and inspecting the results. After removing obviously discrepant data the clean components were returned to the edited file, producing a near-final calibrated data set for that day's observation.

\subsection{Challenges}\label{sec-challenges}

The VLA is currently undergoing the transition to the Expanded VLA (EVLA), with upgrades to the receivers, electronics, and data transmission systems. Along with a new state-of-the-art correlator, due to be commissioned in 2009, the EVLA will provide near continuous frequency coverage from 1 to 50~GHz and much greater observed bandwidths. These will yield a continuum sensitivity improvement ranging from about 5 to over 20 times the current VLA. Observing in the midst of this transition to EVLA did produce some additional challenges. The E-CDF-S observations consisted of 15 to 17  ``classic'' VLA antennas and six to 11 that had been retrofitted into EVLA antennas. In general, the two antenna types performed similarly. More significant than the upgrade of antennas to EVLA compatible ones was the retirement of the original VLA control computers which occurred on 27 June 2007. Slight errors in how the new online system calculated the (u,v,w) coordinates were found, and these errors lead to the distortion of sources. These errors have now been fixed for all the AM889 data residing in the VLA Archive, but we adopted the standard procedure of running UVFIX on all our data as a first step. This task calculates the (u,v,w) data using the antenna positions that are also included in the loaded raw data. Any small residual phase errors resulting from minor errors in the antenna positions are then corrected during the self calibration steps.

Another system glitch reversed the channel indexing for brief periods on three of the observe dates (August 23, September 6, and September 12). We wrote a code using existing AIPS tasks to split out the data with reversed channel indexing, correct its channel order, then return it to the uneffected data. As with the (u,v,w) calculation errors, this problem has now been fixed for all AM889 data in the VLA Archive.

There was also a somewhat mysterious error that has since been attributed to the ``self-test'' procedure within the correlator. This was seen for all VLA observations using a 25~MHz bandwidth, which was the case for the E-CDF-S observations. It amounts to offsets to the raw data corresponding to a given baseline and correlator, which result in ripples around the phase center of the imaged data. These correlator offsets were relatively constant over a five-hour track, although they did vary from day to day. Thus, the AIPS task UVMTH was used to time average the visibilities in each day's (u,v) data set and subtract the averages from the raw data -- thus removing the offsets. The procedure works best when there are no strong sources near the phase center of the affected data, which fortunately was true for the six E-CDF-S pointings (although the second pointing does have a bright source about 4\arcmin{} from the center). Although calibration does affect the correlator offsets, their magnitude was usually small and the gain and phase stability of the observations was very good. This means that the correlator offsets present in the raw (u,v) data remain essentially constant in the calibrated data and hence can safely be removed after calibration. We thereby applied UVMTH in the final steps associated with each individual day's data. First, we took the self-calibrated (u,v) data and subtracted the clean components obtained during the imaging of such data. This produced a (u,v) data set consisting predominantly of the thermal noise and any correlator offsets. UVMTH was run on these data, after which the clean components were added back in to produce the final (u,v) data for that date. This procedure generally resulted in an improvement to the rms noise at the phase center of around 2 $\mu$Jy.

A final challenge that was surmounted involved the data weights. System temperature values for each antenna are included with the data, and the weight assigned any baseline is inversely proportional to the product of the system temperatures for the two antennas that compose that baseline. These weights are calibrated in the same steps that calibrate the data in general. The data for some of the more-recently retrofitted EVLA antennas (Antenna 19 in particular) had incorrect system temperature values, which inadvertantly caused AIPS to overweight these antennas by a factor of three during calibration. This can effectively be thought of as a change in array geometry (packing three antennas into one position in the array) which propagates into the synthesized beam and hence imaging results. We manually adjusted these incorrect weights by the noted factor of three at the conclusion of our reductions for each observing date. AIPS has subsequently been patched to appropriately handle the incorrect system temperature values and hence produce more accurate calibrated data weights.

Finally, as indicated in the previous section some of the data were affected by RFI. When present, this was almost always confined to the higher frequency channels of our observations (around 1450~MHz) and restricted in duration to about one hour. Furthermore, the timing of the onset of the RFI was near daybreak. When editing data we usually excised the highest frequency channels during the effected time range, and attempted to keep the remaining data. Additional removal of data compromised by RFI occurred after imaging and self calibration during the step where clean components of imaged sources were removed and the resulting (u,v)-subtracted data were edited.

\subsection{Combining Data}

Once the data for each individual observation date had been satisfactorily calibrated and edited, they were combined. All eight (in some cases, 9) data sets corresponding to a given pointing had their times converted to hour angles, at which point they were concatenated and then averaged by baseline. For the September 23 data, which were obtained after the VLA had moved into its BnA configuration, we only included those antenna locations which were present for the preceding observations. The BnA array has the same antenna configuration for its north arm as the A array, with the east and west arms moved to a more compact arrangement of antennas. Thus, for the September 23 data only 17 antennas were used (the nine for the north arm, with four from each the east and west arms). The resulting concatenated and baseline averaged data were then imaged as in the procedure for individual observing tracks. For this imaging, we repeated the step of making a large low-resolution image to identify flanking fields since the combined data were deeper and hence revealed additional outlying sources. We also expanded the fly's-eye by an additional ring making it consist of 127 facets thus approximating a 30\arcmin{} radius circle and hence the full VLA primary beam. In total, about 165 facets were imaged at the full resolution for each pointing. These images were inspected and sources boxed, with many additional sources being identified relative to the individual days' data on account of the $\sim\sqrt{8}$ improvement in sensitivity. 

A subsequent round of imaging at the full resolution, using the more complete listing of boxed sources, was then used as the input source model for a final self calibration. In this case, the self calibration adjusted both amplitude and phase. This is particularly useful when combining data collected on different days, as it puts the individual gains onto a common system. The self-calibrated data were then imaged to produce the final maps corresponding to that pointing in the mosaic. The rms noise in the central facet of each pointing ranged from about 10.3 to 11.1 $\mu$Jy beam$^{-1}$. In each case, the source-subtracted residual images were well approximated by a Gaussian noise distribution with maxima and minima usually in the $\pm55$ $\mu$Jy range. Slightly larger values did appear in facets containing the brighter sources.

The six pointings were then stitched together using the AIPS task FLATN to produce the final mosaic image that comprises the first data release. FLATN properly weights the data by the inverse square of the primary beam power pattern, with net weights resulting from the summed contributions of up to six separate pointings. Note that regions within a given pointing that were included in multiple imaged facets of that pointing do not receive extra weight. For each pointing we included data up to the one-third power point of the primary beam (about 19\arcmin; refer to Figure \ref{fig-flyseye}), noting the tradeoff of sensitivity with bandwidth and time-averaging smearing. The output mosaic maintains the 0\farcs5 pixel size but increases the overall image size to $4096 \times 4096$ centered on 3$^{\mbox{{\scriptsize h}}}$32$^{\mbox{{\scriptsize m}}}$28\fs0 $-$27$^\circ$48\arcmin30\farcs0. Thus, the final image is about 34\farcm1 $\times$ 34\farcm1 and extends to just beyond the edges of the formal E-CDF-S. It is available for download upon request from the author.

\begin{figure}
\figurenum{4}
\epsscale{0.9}
\plotone{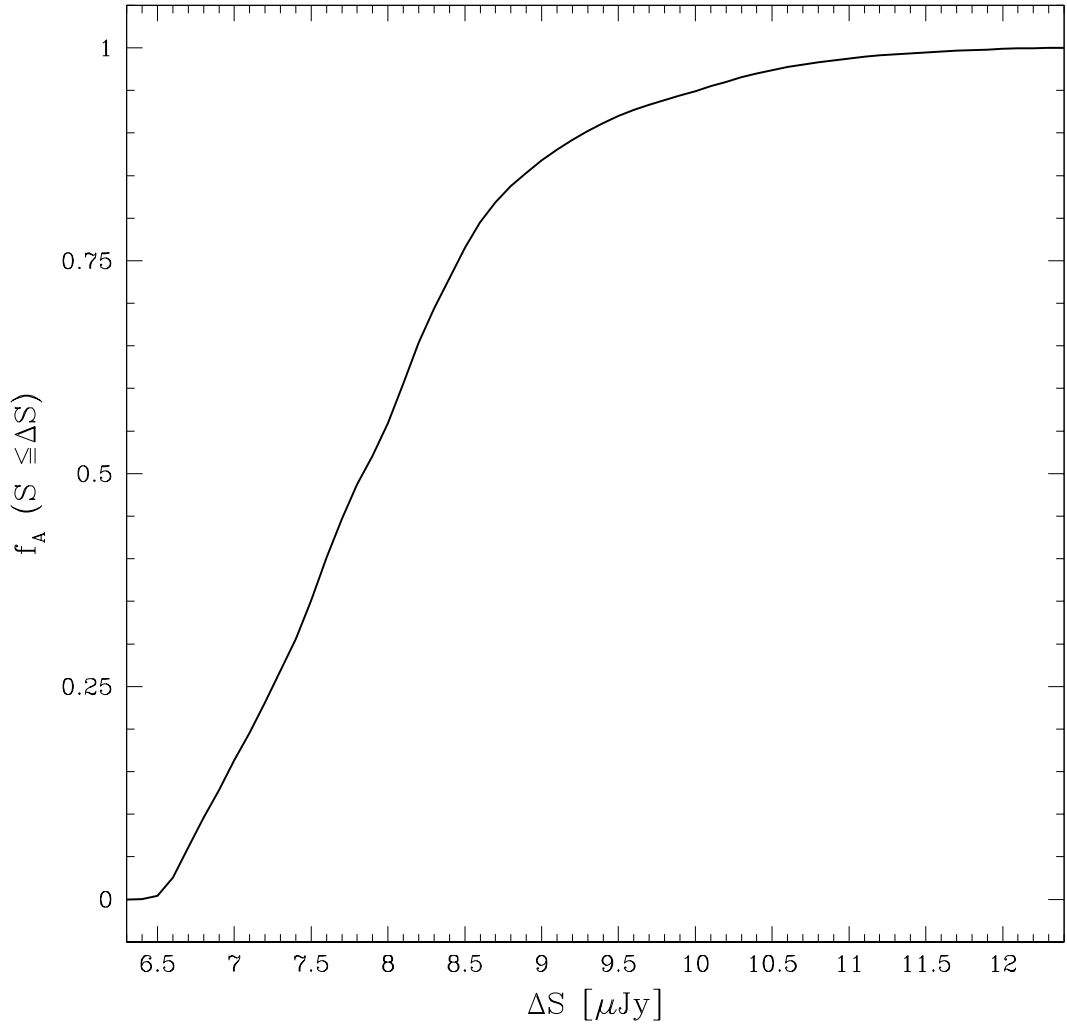}
\caption{Fractional area covered by the first data release mosaic image at a given rms sensitivity or better. This plot corresponds to the central $32\arcmin \times 32\arcmin$ (the approximate E-CDF-S X-ray coverage area) of the full mosaic image.\label{fig-rmsfrac}}
\end{figure}

\begin{figure}
\figurenum{5}
\epsscale{0.9}
\plotone{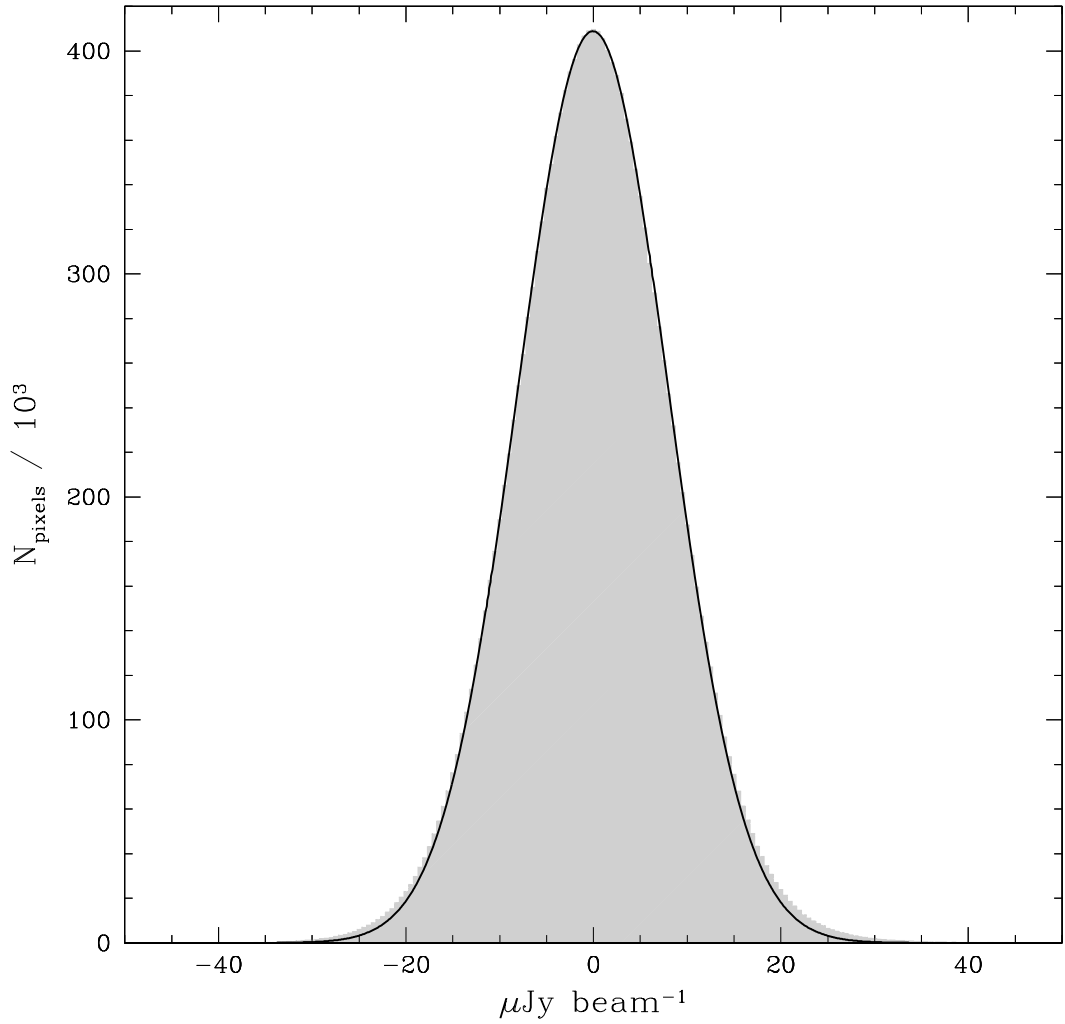}
\caption{Noise histogram measured over the full first data release mosaic image. The noise is well fit by a Gaussian with $\sigma=8.0 \mu$Jy, shown as a solid line.\label{fig-noisehist}}
\end{figure}

The characteristics of the map were determined by evaluating the local rms noise at every location. This was achieved using a 2\arcmin{} box size, and contours of the resulting rms map are shown in Figure \ref{fig-rms}. It can be seen that the local noise is at its lowest in the center and increases roughly radially, with some variation caused by the slightly increased noise level around the brighter sources (refer to Figure \ref{fig-msc}). The fraction of the area covered at a given rms sensitivity or better is shown graphically in Figure \ref{fig-rmsfrac}. This figure also shows that the most sensitive regions have local rms noise near 6.4 $\mu$Jy beam$^{-1}$ while effectively all of the image has an rms of 12 $\mu$Jy beam$^{-1}$ or better. Within the central $32^\prime \times 32^\prime$ (i.e., the E-CDF-S region) the rms noise is 7.9 $\mu$Jy beam$^{-1}$. As for the individual pointings, the noise in the mosaic image is fairly well described by a Gaussian (Figure \ref{fig-noisehist}).

\section{Source Catalog}\label{sec-catalog}

\subsection{Generation of Catalog}

To generate the source catalog, we constructed a signal-to-noise ($S/N$) image by dividing the final mosaic image by the computed local rms image. Sources within the $S/N$ image were identified and fitted by Gaussians using the AIPS task SAD, with the task instructed to reject sources with peaks less than 5$\sigma$ and produce an output list of all accepted sources. SAD also created a residual image consisting of the input $S/N$ image minus the Gaussian fits to accepted sources. The residual map was then inspected to identify missed sources as well as accepted sources which were poorly fit by SAD. In some cases these included single sources which were split into two sources each with lower peak flux density. Missed sources were added to the preliminary source list, and poorly fit sources were flagged for later follow-up. 

This modified source list was than used as input to produce the final source catalog. All sources with peak flux density greater than seven times the local rms noise were fit using the task JMFIT, which like SAD fits Gaussians to sources individually specified by the user. The output coordinates, peak flux density and error, integrated flux density and error, and a resolution index were recorded. Astrometric observations with the VLA at 1.4~GHz indicate a relative positional accuracy of better than 0\farcs1, with additional errors in source coordinates arising from source fitting. This latter error implies that catalog sources near the detection threshhold will have positional errors of up to 0\farcs2. JMFIT uses the rms noise of the overall image for determining the errors in the peak and integrated flux densities, so we adjusted these quantities using the local rms noise taken directly from the rms noise map at the coordinates of the source. This measurement was adopted as the error in the peak flux density, and the error in the integrated flux density was taken to be the JMFIT value multiplied by the ratio of the local rms noise divided by the rms noise assumed by JMFIT. We relied on the JMFIT deconvolved major and minor axes of the sources in evaluating whether a source might be resolved. In addition to the nominal deconvolved size of each axis, JMFIT provides a estimates of their maximum and minimum deconvolved sizes. Sources for which the minima for both the deconvolved major and minor axes were zero, and hence consistent with being unresolved, were assigned a resolution index value of zero. Sources which were apparently resolved received a one for each axis having a non-zero minimum. Thus a source that JMFIT indicated was resolved on both its major and minor axes was assigned a value of two. In this scheme, all sources with non-zero resolution index are considered resolved. Sources that were poorly fit by Gaussians during the running of SAD and subsequent inspection were evaluated using the task TVSTAT, which allows the user to interactively set an irregularly-shaped aperture around the source and then returns the number of pixels in the region, their mean value, the coordinates corresponding to the maximum flux density, and the integrated flux density. By definition these sources were resolved and were denoted by using a three for their resolution index in the catalog. The errors in their integrated flux densities were determined as the square-root of the number of beams contained by the aperture times the local rms.

We note here and in the next section that some caution must be observed when considering whether a source is truly resolved, and subsequently whether to adopt the peak or integrated value for the flux density of that source. In addition to those considerations generic to Gaussian source fitting of radio images \citep[e.g.,][]{condon1997}, the present image is the mosaic of six individual pointings. This means that every individual source can have contributions from up to six separate pointings, each of which suffers some degree of bandwidth smearing. The total flux density of a source will be conserved, although the peak flux density is likely to be lower than its true value on account of bandwidth smearing. Consequently, for each source that JMFIT suggested was resolved in the mosaic image we have examined the six images corresponding to the individual pointings. Although these will be of lower signal-to-noise than the final mosaic, the effect of bandwidth smearing on sources within these individual pointings can be assessed using JMFIT. We thus noted which sources that appeared resolved in the mosaic image but were unresolved in the pointing that had its center nearest the source in question. For these sources, we maintained the resolution index obtained from the fit to the mosaic image but changed its sign. This preserved the resolution information determined from the mosaic image, but identified sources where effects such as bandwidth smearing might cause an intrinsically unresolved source to appear resolved.

We have chosen to include a 7$\sigma$ catalog with this release instead of a lower threshold catalog for several reasons. The primary one is that this is a first data release, and we will produce a deeper image and more carefully constructed source catalog for the next data release (see Section \ref{sec-future}). Providing a 7$\sigma$ catalog provides some assurance that the parameters of identified sources will not change significantly in the next data release and that the incidence of spurious sources will be minimal. In addition, the 7$\sigma$ catalog is a nice match to the source catalog of K2008, as discussed in the next section (\ref{sec-comparisons}). Finally, the FITS image corresponding to this first data release is available along with the catalog, thereby allowing interested researchers the opportunity to determine their own source lists and properties.

The catalog is presented in Table \ref{tbl-catalog}. It is organized by increasing right ascension, and includes the $S/N$ of each source (defined as the fitted peak divided by the local rms), peak flux density and error, integrated flux density and error, an index evaluating whether the source is resolved (see above), and cross referencing of the sources to other radio surveys of the E-CDF-S \citep[K2008, A2006, ][]{norris2006}. In fact, we have relied on the classifications of the K2008 survey for multiple component sources. Thus, the individual components associated with features arising from a single source are grouped together (for example, the core and lobes of a powerful radio galaxy). Table \ref{tbl-catalog} consists of 464 separate sources, 19 of which are broken down into multiple components. The position and integrated flux density of the multiple component sources are presented in the table, followed by entries for their individual components with positions indented for clarity. One source with $S/N = 6.4$ was included in Table \ref{tbl-catalog} because it was clearly resolved and had an integrated flux density nearly six times greater than its associated error (the coordinates for this source are 3$^{\mbox{{\scriptsize h}}}$32$^{\mbox{{\scriptsize m}}}$35\fs01 $-$27$^\circ$55\arcmin32\farcs8). This appears to be the only resolved source for which $S/N < 7$ yet the integrated flux density is significant at greater than 5$\sigma$.

\subsection{Comparison to Prior Surveys}\label{sec-comparisons}

The K2008 survey provides a check on the source list and parameters of Table \ref{tbl-catalog}. There are 266 sources in K2008 (including some with multiple components), 240 of which are in Table \ref{tbl-catalog} (including the one exception for which $S/N < 7$). The majority of the K2008 sources not included in Table \ref{tbl-catalog} (21/26) are detected in the first data release mosaic image with $5 \leq S/N < 7$. Four of the remaining five K2008 sources not included in Table \ref{tbl-catalog} also are present in the first data release mosaic image with $3 \leq S/N < 5$, meaning source \#6 of K2008 is the only object which does not seem to be detected in our observations. This source is a 5.1$\sigma$ detection in K2008, with a flux density of $110\pm23$ $\mu$Jy and a position of 3$^{\mbox{{\scriptsize h}}}$31$^{\mbox{{\scriptsize m}}}$14\fs87 $-$27$^\circ$55\arcmin43\farcs4. The maximum detected peak flux density within 5$^{\prime\prime}$ of this position in the E-CDF-S first data release mosaic is 16.5 $\mu$Jy (about 1.7$\sigma$).

\begin{figure}
\figurenum{6}
\epsscale{0.9}
\plotone{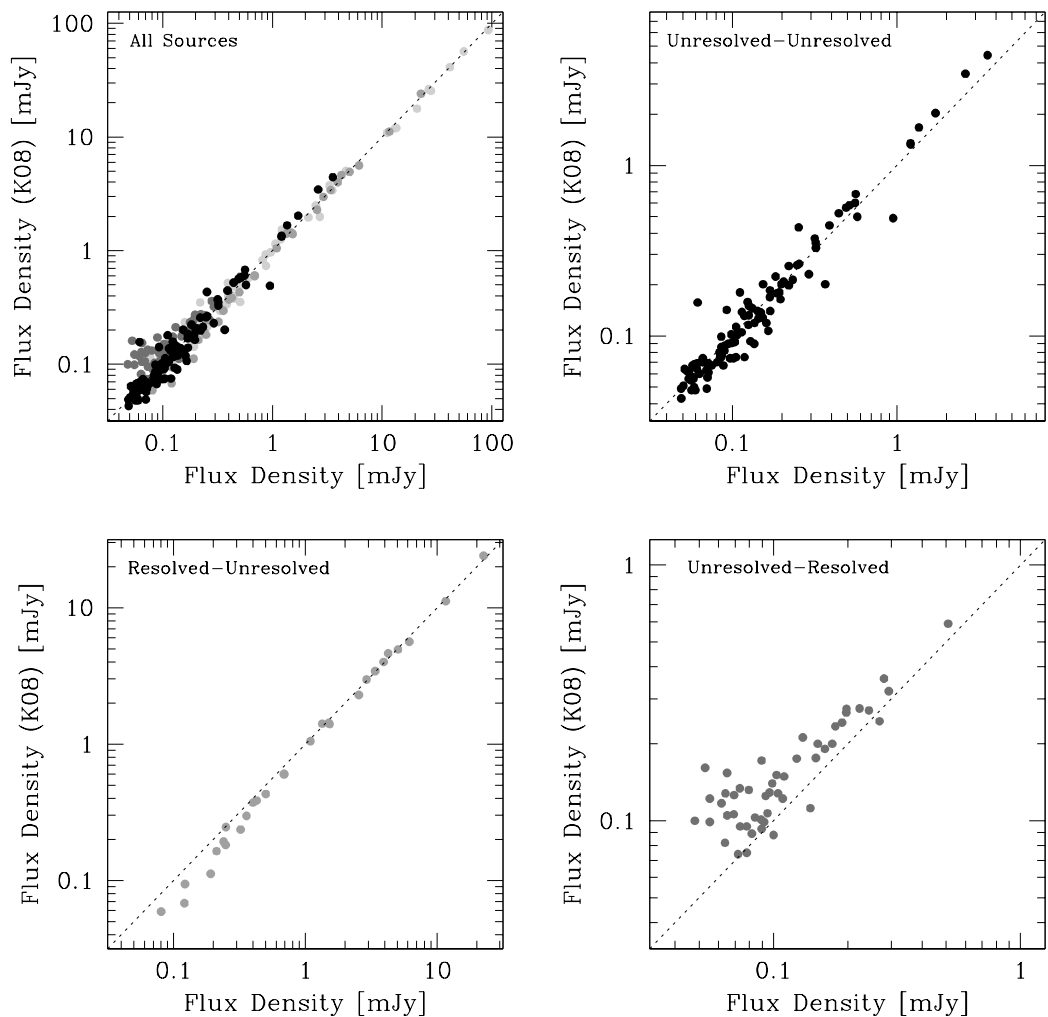}
\caption{Comparison of new fluxes with \citet{kellermann2008}. The top left box includes all sources color-coded by source morphology, where black points indicate sources which are unresolved in both catalogs (plotted alone in the top right box), light grey points are for sources resolved in both catalogs, medium grey points are for sources resolved by the present survey yet unresolved in \citet{kellermann2008} (bottom left box), and dark grey points are for sources unresolved in the present survey yet resolved in \citet{kellermann2008} (bottom right box).\label{fig-fluxcomp}}
\end{figure}

We have adopted the peak flux density for unresolved sources and the integrated flux density for resolved sources (i.e., those with resolution index greater than zero) as the total flux density of that source \citep[e.g.,][]{owen2005}, and compared these values with the corresponding flux densities in K2008. In the case of sources separated into multiple components, we compared only the total measured flux densities in the two surveys. It can be seen from Figure \ref{fig-fluxcomp} that the source flux densities in the two surveys are broadly consistent. Sources that differ from a one-to-one correspondence may result from intrinsic differences in the flux density calibration of the two surveys, the slightly different resolutions of the surveys (2\farcs8$\times$1\farcs6 compared to the 3\farcs5$\times$3\farcs5 of K2008), and real source variability. The top right panel of Figure \ref{fig-fluxcomp} depicts only those sources which are unresolved in both surveys, as these are the most direct comparison of derived flux densities. A slight scaling difference may exist in the sense that the Table \ref{tbl-catalog} flux densities for these unresolved sources are $\sim3\%$ greater than the K2008 values, although the significance of this difference is only 1.6$\sigma$. This was determined after applying 3$\sigma$ clipping to remove outliers such as variable sources and mis-matched sources, which in this case amounted to three sources. Similarly, comparison of the flux densities for sources which were resolved in both surveys indicated the flux densities were consistent, with the measurements from the current survey being higher than the K2008 values at a significance of only 0.4$\sigma$.

Some unresolved sources of K2008 had Gaussian fits in our data that suggested the sources were resolved, and these are shown in the bottom left panel of Figure \ref{fig-fluxcomp}. Similarly, some resolved sources in K2008 were apparently unresolved in our data. This seemingly counter-intuitive characterization might result from sources consisting of an unresolved core plus diffuse and faint emission, for example. Sources of this type are shown in the bottom right of Figure \ref{fig-fluxcomp}, and follow the same general pattern as those resolved in the current data but unresolved in K2008: the measured flux density coming from the survey in which the source is apparently resolved exceeds that from the survey in which it appears unresolved. Presumably this is the result of uncertainties in Gaussian fitting to faint sources in the presence of noise \citep{condon1997}, although it is also consistent with the expectations of source confusion (one survey may separate nearby sources while the other combines them into a single extended source).

Additionally, some of the possible slight differences in flux densities may result from different criteria used to assess whether a source is resolved. K2008 examined the difference between the peak and integrated flux density, and if this difference was less than twice the local rms noise the source was considered unresolved. For such sources the flux density was taken to be the average of the peak and integrated measurements from JMFIT. We have applied this definition and repeated our tests comparing flux densities for sources in the present survey with those of K2008. As with comparisons made using the resolution index included in Table \ref{tbl-catalog}, the flux densities for sources resolved in both surveys were consistent. Sources which were unresolved in both surveys had a slight tendency for the present survey to have higher flux densities (about $8\%$, with a 2.6$\sigma$ significance that the flux densities are greater than those in K2008). Other procedures commonly used in the literature to evaluate whether a source is resolved also compare peak and integrated flux densities. For example, \citet{huynh2005} applied such a scheme to deep 1.4~GHz observations of the Hubble Deep Field - South. They noted that cases where the integrated flux density is less than the peak flux density can safely be assumed to be unresolved sources where the differing flux density measurements are the result of noise in the image and its effect on source fitting. Since this noise can be expected to result in equal numbers of unresolved sources with peak flux density greater than integrated flux density to those with peak flux density less than integrated flux density, the former can be used to set a threshold (usually as a function of signal-to-noise) for defining a source as resolved. Our inclusion of both peak and integrated flux densities in Table \ref{tbl-catalog}, along with making the first data release mosaic image publicly available, provides interested researchers with the information required to apply whichever definition of resolved and unresolved sources they see fit. We note here that using the resolution index as described in Section \ref{sec-catalog} and included in Table \ref{tbl-catalog} provides a good match to the K2008 results if one adopts peak flux densities for unresolved sources and integrated flux densities for resolved ones.

Table \ref{tbl-catalog} also includes entries for 51 of the 64 sources in the A2006 ATCA survey data coincident with the GOODS field. As with the K2008 comparison, many of the A2006 sources not present in Table \ref{tbl-catalog} are detected in the first data release mosaic image with $S/N < 7$. However, three of the fainter A2006 sources are not detected (i.e., the maximum in the $S/N$ image within 5\arcsec{} of the A2006 position is less than two): A2006 \#3, \#39, and \#40. These are among the more uncertain detections in A2006, with flux densities of $87\pm42$, $82\pm25$, and $63\pm24$ $\mu$Jy, respectively. Two further A2006 sources coincide with low significance detections in the first data release mosaic image, although with slight positional offsets (A2006 \#45 and \#50). Comparison of the flux densities of sources in common again indicates a possible offset in the flux calibration of the present survey, with our flux densities being $\sim11\%$ greater than the A2006 values. This is a 2.6$\sigma$ result based on all common sources regardless of whether they are resolved, as such information is not present in A2006. We confirm the result noted in the \citet{norris2006} analysis of the ATCA data and discussed further in K2008, in that the flux density scale of the \citet{norris2006} catalog appears low by $\sim20\%$ relative to other surveys.

In summary, there is good agreement between Table \ref{tbl-catalog} and prior 1.4~GHz data for the E-CDF-S. The majority of sources detected in K2008 and A2006 are also detected in the first data release mosaic image, with most of these having $S/N > 7$ and hence being included in Table \ref{tbl-catalog}. We find slight evidence for an offset in the flux calibration scale, with the Table \ref{tbl-catalog} values on average appearing greater than those in K2008. The magnitude of this scale factor depends on the adopted definition for a resolved source, and amounts to about 3\% if one adopts the resolution index and peak flux densities from Table \ref{tbl-catalog} and compares sources that were unresolved by both studies.

\section{Future Improvements}\label{sec-future}

Although this first data release achieves good rms sensitivity across the full E-CDF-S area, a number of improvements to the data reduction are possible. As indicated in Section \ref{sec-challenges}, there were several glitches with the raw (u,v,w) data which were overcome by various procedures. These have now been corrected for all data in the VLA Archive, meaning downloads of the raw data and reductions ``from scratch'' will avoid some of the forced fixes that were implemented in our reductions, along with any downstream effects thereof. Re-doing the reductions on several individual days of data starting with corrected data from the VLA Archive suggest an improvement of $\sim5\%$ in the achieved rms sensitivity.

In addition, there are several well-known and time-intensive reduction procedures that we have not yet implemented but will implement for the next data release. Most of these amount to imaging the data in segments rather than all together. First, the VLA right circular polarization and left circular polarization beams are slightly separated, meaning they have slightly different pointings on the sky (``beam squint''). Imaging the two polarizations separately and then optimally combining the output maps can therefore improve the noise level and dynamic range of the final map. As with most imaging considerations at 1.4~GHz the effect is greatest for sources distant from the phase center, the sidelobes of which degrade the overall image. Similarly, observing in two bands at separate intermediate frequencies straddling 1.4~GHz (1.365~GHz and 1.435~GHz) results in slightly different resolutions and consequently slightly different instrumental gains. The solution is the same as for the separate polarizations, in that the two intermediate frequencies can be imaged separately, restored to the same resolution, and then combined. Next, over the course of each five-hour observing track the shape of the primary beam rotates on the sky and sources away from the pointing center appear to vary. Again, this can be alleviated by imaging the data in smaller portions set by fixed ranges of observed hour angle and combining the resulting images. Finally, new functionality in AIPS provides improved self-calibration of bright interfering sources (``PEELR''). This will clean up much of the ``ripples'' around bright sources, noticeable in Figure \ref{fig-rms} as slight increases in the local rms sensitivity around such sources. From experience with other programs, we anticipate the net improvement resulting from the combination of these improved imaging steps to be on the order of 10\%.

The next data release will also include improvements to the source catalog. It will increase the total number of sources on account of improvements to the sensitivity and application of a less conservative detection limit. As implied in Section \ref{sec-comparisons} there are hundreds of likely real detected sources with peak flux densities in the $5\sigma - 7\sigma$ range. The characteristics of all detected sources will also be more rigorously determined. In Table \ref{tbl-catalog} we have generally assumed that the Gaussian fit parameters determined by JMFIT were a good representation of any given source. It is known, however, that Gaussian fits to faint sources in the presence of noise are biased and yield overestimates for peak flux densities \citep[e.g.,][]{condon1997}. Comparison of flux densities measured through Gaussian fits with ones derived through photometry performed in concentric apertures can be used to produce more accurate flux densities \citep[e.g., see][]{simpson2006}. Perhaps the most glaring omission from the information in Table \ref{tbl-catalog} is the lack of information on source size other than a descriptive index on whether the source appeared to be resolved. Better handling of the combined effects of bandwidth and time average smearing from six separate pointing centers is needed in order to ascertain whether sources are truly resolved or merely appear to be due to instrumental effects. This will also allow for the reporting of reasonable estimates and associated errors for source sizes.

\section{Summary}\label{sec-summary}

We have presented the survey strategy, observational details, data reduction procedure, and first data release for the VLA 1.4~GHz survey of the Extended {\it Chandra} Deep Field South. The mosaic image corresponding to this first data release contains the full E-CDF-S area and is available by contacting the author. It reaches an rms sensitivity of 6.4 $\mu$Jy per 2\farcs8$\times$1\farcs6 beam at its most sensitive point, with a typical rms sensitivity of 8 $\mu$Jy across the full image. A 7$\sigma$ catalog of 464 sources, 19 of which consist of multiple components, is provided in Table \ref{tbl-catalog}. Comparison with existing surveys confirms that the relative flux density scale is accurate to within a few percent, although users are cautioned that source extraction for fainter sources leads to larger errors.

The importance of the E-CDF-S to the astronomical community has driven us to provide this first data release in an expeditious manner. Significant improvements in the achievable rms sensitivity and catalog construction are possible, and we anticipate a second data release including these improvements for January 2009.

\acknowledgments
We gratefully acknowledge the hard work of the NRAO staff. That this program was successful in the midst of a very dynamic time in the transition to EVLA is a strong testament to their dedication. We also thank Mark Dickinson, Anton Koekemoer, Rob Ivison, and Glenn Morrison for valuable discussions about the motivations and planning of this program, and the anonymous referee for prompting additional testing to confirm the reliability of the reported flux densities.

\clearpage

\begin{deluxetable}{l l l l}
\tablecolumns{4}
\tablecaption{Observation Summary\label{tbl-obs}}
\tablewidth{0pt}
\tablehead{
\colhead{Pointing ID} & \colhead{RA(J2000)} & \colhead{Dec(J2000)} & \colhead{Dates Observed}}
\startdata
ECDFS1A\tablenotemark{a} & 03:33:22.25 & -27:47:30.0 & Jun 03 \\
ECDFS1  & 03:33:22.25 & -27:48:30.0 & Jul 12, Jul 19, Jul 30, Aug 03\tablenotemark{b}, Aug 11\tablenotemark{b}, \\
        &             &             & Aug 21\tablenotemark{b}, Sep 02\tablenotemark{c}, Sep 11 \\
ECDFS2  & 03:32:55.12 & -27:38:03.0 & Jun 15, Jul 13, Jul 20\tablenotemark{b}, Jul 27, Aug 04, \\
        &             &             & Aug 13\tablenotemark{b}, Aug 23\tablenotemark{b}, Sep 06\tablenotemark{b}, Sep 23\tablenotemark{d} \\
ECDFS3  & 03:32:00.88 & -27:38:03.0 & Jun 24, Jul 14, Jul 21, Jul 28, Aug 05, \\
        &             &             & Aug 14\tablenotemark{b}, Aug 25\tablenotemark{b}, Sep 07 \\
ECDFS4  & 03:31:33.75 & -27:48:30.0 & Jun 25, Jul 15, Jul 22, Jul 29, Aug 06\tablenotemark{b}, \\
\enddata

\tablenotetext{a}{Erroneously offset by 1\arcmin{} in declination relative to intended mosaic.}
\tablenotetext{b}{Radio Frequency Interference is present, usually for about one hour and affecting higher frequency channels.}
\tablenotetext{c}{Data were lost due to power outage at VLA site.}
\tablenotetext{d}{Data collected in BnA configuration.}
\tablenotetext{e}{Data were lost due to complications related to EVLA transition.}
\tablenotetext{f}{Scheduled power outage at VLA site during middle of Aug 17 track. Remainder of 5-hour track observed on Sep 12.}
\tablecomments{All observations occurred in 2007 and consisted of five-hour tracks centered on 0330 LST unless otherwise noted.}

\end{deluxetable}

\begin{deluxetable}{l l r r r r r r r r}
\tablecolumns{10}
\tablecaption{Preliminary Source List\label{tbl-catalog}}
\tablewidth{0pt}
\tablehead{
\colhead{RA} & \colhead{Dec} & \colhead{SNR} & \colhead{$S_p$} & \colhead{$\Delta S_p$} &
\colhead{$S_i$} & \colhead{$\Delta S_i$} & \colhead{Res} & \colhead{VLA ID} & \colhead{ATCA ID}\\
\colhead{(J2000)} & \colhead{(J2000)} & \colhead{} & \colhead{[$\mu$Jy]} & \colhead{[$\mu$Jy]} &
\colhead{[$\mu$Jy]} & \colhead{[$\mu$Jy]} & \colhead{} & \colhead{} & \colhead{}\\
\colhead{(1)} & \colhead{(2)} & \colhead{(3)} & \colhead{(4)} & \colhead{(5)} & \colhead{(6)} &
\colhead{(7)} & \colhead{(8)} & \colhead{(9)} & \colhead{(10)} 
}
\startdata
03 31 10.69 & -28 03 22.9 &    17.4 &    207.1 &    12.1 &    350.0 &  30.4 &  2 & \nodata & C335 \\
03 31 10.81 & -27 55 52.8 &    12.4 &    120.8 &     9.7 &    152.7 &  19.8 & -1 & \nodata & \nodata \\
03 31 10.94 & -27 58 10.5 &    12.4 &    125.9 &    10.1 &    155.2 &  20.1 & -1 & \nodata & \nodata \\
03 31 11.48 & -27 52 59.0 &    11.8 &    109.2 &     9.0 &    156.0 &  20.1 &  0 & \nodata & C337 \\
03 31 11.69 & -27 31 44.2 &   205.  &   2607.  &    13.1 &   3391.  &  56.6 &  3 &       1 & C339 \\
03 31 11.82 & -27 58 17.9 &     8.0 &     82.3 &    10.2 &    138.9 &  25.5 &  2 & \nodata & \nodata \\
03 31 12.16 & -28 00 49.7 &     8.3 &     91.3 &    11.0 &     89.7 &  18.8 &  0 & \nodata & \nodata \\
03 31 12.60 & -27 57 18.3 &    15.0 &    148.8 &     9.9 &    154.6 &  17.5 &  0 & \nodata & \nodata \\
\enddata

\tablenotetext{a}{Although this source has $S/N < 7$ for its peak flux density, its integrated flux is well over a 5$\sigma$ detection and hence it has been included in this catalog.} 

\tablecomments{Column descriptions: (1) \& (2) Right Ascension and Declination of radio source (J2000); (3) signal-to-noise ratio determined as source peak flux density over local rms; (4) and (5) peak flux density and error; (6) \& (7) integrated flux density and error; (8) resolution index, defined as 0 if source is unresolved on both axes in mosaic image, 1 if resolved on one axis only, 2 if resolved on both axes, and 3 if resolved on both axes and the integrated flux density was measured through aperture photometry rather than Gaussian fitting. When preceded by a negative sign, the source was unresolved in the individual pointing with the nearest field center; (9) ID from \citet{kellermann2008}, where capital letters indicate components of a larger source and lower case letters indicate new components resolved by the current observations; (10) ID from the ATCA surveys, where ``A'' indicates \citet{afonso2006} and ``C'' are the component IDs from \citet{norris2006}.}

\end{deluxetable}

\clearpage

\end{document}